\documentclass[a4paper,11pt,dvips]{article}
\usepackage{graphicx}
\usepackage{amsmath}
\usepackage{amssymb}
\usepackage{latexsym}
\usepackage{color}
\newcommand{\bra}{\begin{array}}
\newcommand{\era}{\end{array}}
\newcommand{\beq}{\begin{equation}}
\newcommand{\eeq}{\end{equation}}
\newcommand{\beqar}{\begin{eqnarray}}
\newcommand{\eeqar}{\end{eqnarray}}

\def\BC{\bb C}
\def\_\BC{\bbi C}

\def\( {\left(}
   \def\) {\right)}
\def\[ {\left[}
\def\] {\right]}
\def\no2 {{\textstyle{n\over 2}}}

\begin{document}
\begin{titlepage}
\setcounter{page}{1}
\renewcommand{\thefootnote}{\fnsymbol{footnote}}

\begin{flushright}
ucd-tpg:1103.03\\
%arXiv:yymm.xxxx
\end{flushright}

\vspace{5mm}
\begin{center}
{\Large \bf {Solution of One-dimensional Dirac Equation via Poincar\'e Map }}

\vspace{5mm}

{\bf Hocine Bahlouli}$^{a,b}$, {\bf El Bou\^azzaoui Choubabi}$^{a,c}$
and {\bf Ahmed Jellal$^{a,c,d}$\footnote{\sf ajellal@ictp.it -- ahjellal@kfu.edu.sa}}

\vspace{5mm}

{$^a$\em Saudi Center for Theoretical Physics, Dhahran, Saudi Arabia}

{$^b$\em Physics Department,  King Fahd University
of Petroleum $\&$ Minerals,\\
Dhahran 31261, Saudi Arabia}

{$^{c}$\em Theoretical Physics Group,  %Department of Physics,
Faculty of Sciences, Choua\"ib Doukkali University},\\
{\em PO Box 20, 24000 El Jadida,
Morocco}

{$^d$\em Physics Department, College of Sciences, King Faisal University,\\
PO Box 380, Alahsa 31982, Saudi Arabia}

%{$^{d}$\em Max Planck Institute for the Physics of Complex Systems,\\
%N\"othnitzer Str. 38, 01187 Dresden} %\\[1em]

%\vspace{30mm}

\vspace{3cm}

\begin{abstract}
We solve the general one-dimensional Dirac equation using a "Poincar\'e Map" approach which avoids any
approximation to the spacial derivatives and reduces the problem to a simple recursive relation
which is very practical from the numerical implementation point of view. To test the efficiency
and rapid convergence of this approach we apply it to a vector coupling Woods--Saxon potential,
which is exactly solvable.
Comparison with available analytical results is impressive and hence validates the accuracy and
efficiency of this method.

\vspace{3cm}

\noindent PACS numbers: 73.63.-b; 73.23.-b; 11.80.-m

\noindent Keywords: Dirac equation, mapping, Woods--Saxon, numerical.

\end{abstract}
\end{center}
\end{titlepage}

The Dirac equation describes a relativistic particle of spin one-half at high velocities
(below the threshold of pair production)~\cite{1}. It describes the state of electrons in
a way consistent with special relativity, requiring that electrons have spin $1/2$, and predicting
the existence of an antiparticle partner to the electron (the positron). The physics and mathematics
of the Dirac equation is very rich, it is a first order matrix linear differential equation whose solution
is a 4-component wavefunction (a spinor). Nevertheless, it was hard to find nontrivial exact solutions of
this equation. Until 1989, there was only one nontrivial exact solution of the Dirac equation for the very
important Coulomb problem. By convention, a nontrivial exact solution of the Dirac equation is a solution that,
in the non-relativistic limit, reproduces a known solution of Schr\"odinger equation with that specific potential
and hence carries the same relativistic potential name.

In this regard, Moshinsky and Szczepaniak \cite{2} in 1989
were able to formulate and solve the relativistic oscillator problem (Dirac-Oscillator). During the last decade,
Alhaidari has introduced an effective approach to the solution of the Dirac equation for spherically symmetric
potentials \cite{3,4,5,6,7}. His method was initiated by the observation that different potentials can be grouped
into classes; for example, the non-relativistic Coulomb, oscillator and S-wave Morse problems constitute one such class.
Therefore, the solution of two problems in one class implies solution for the remaining one. By this method, the S-wave
Dirac-Morse problem was formulated and solved \cite{3}.

    Using the above-mentioned method, other potentials were treated; among these are Dirac-Scarf,
Dirac-Rosen-Morse I \& II,  Dirac-Poschl-Teller, Dirac-Eckart \cite{4}, Dirac-Hulthen, and Dirac-Woods-Saxon potentials \cite{8}. On the other hand alternative methods such as quasi-exactly solvable problems at rest mass energies with power-law relativistic potentials were investigated following the same procedure \cite{5}. Orthogonal polynomials were also used to find series solutions of Dirac equation for scattering and bound states \cite{6}.
    In this work we will be interested in solving the one-dimensional (1D) Dirac equation in general using a Poincar\'e map which enable
us to treat exactly the spacial derivative operator while the only approximation is made upon discretizing the scattering potential. The approach is very efficient and converges rapidly. To exhibit the efficiency of our method we use it to solve the vector coupling Woods-Saxon potential and compare our results for the transmission coefficient with the analytic one. The accuracy and easy implementation of our method is impressive which suggest its suitability in dealing with 1D Dirac equation for an arbitrary potential. %and its potential extension to two dimensions.

We consider a one dimensional Dirac equation for a particle of mass $m$, subject to a vector potential coupling $V(x)$. In unit system $\hbar = c = 1$,
this particle is governed by the Dirac Hamiltonian %the following Dirac Hamiltonian (in unit system $\hbar = c = 1$)
\begin{equation} \label{GrindEQ1}
H= p_{x}\sigma_{x}+ m\sigma_{z}+ V(x)
\end{equation}
with $\sigma_{x}$ and $\sigma_{z}$ being the Pauli matrices. The stationary eigenvalue equation can be written explicitly in the following form
\begin{equation} \label{GrindEQ2}
\left(
  \begin{array}{cc}
    m+V(x)-\varepsilon & -i \frac{d}{dx} \\
    -i \frac{d}{dx} & -m+V(x)-\varepsilon \\
  \end{array}
\right)\left(
         \begin{array}{c}
           \psi^1 \\
            \psi^2\\
         \end{array}
       \right)=0
\end{equation}
 where $\psi(x)=(\psi^1~~\psi^2)^t$ is the eigenspinors of two components.
Generally speaking for an arbitrary potential $V(x)$, this equation is difficult to solve analytically except for a very limited number of solvable potential as classified by supersymmetric quantum mechanics \cite{11}. Alternatively, in this work an iterative method that was used a lot for 1D Schrodinger equation, the Poincar\'e Map approach \cite{12}, will be adapted to our relativistic problem. Using this approach we are going to show that the above 1D Dirac equation can be replaced by a Poinccar\'e map associated with the above wave equation. This approach will enable us to solve the Dirac equation and generate the spinor wavefunction iteratively, a method very suitable for numerical implementations. In addition, we will be able to use the transfer matrix approach to compute the transmission coefficient.

We consider our one-dimensional system where the  particle is moving under the action of a scattering potential $V(x)$
bound to a region of size $L$ of our system. So our space can be decomposed into three major regions: the extremes are free like regions where the potential is almost zero and an intermediate region where the potential is $V (x)$. Now we proceed in subdividing the potential
interval $L$ into $N+1$ equal regions, in every region we approximate the potential by a constant local value $V_n = V(x_n)$  where $x_n=nh$ and $h=\frac{L}{N+1}$, with $N$ is the maximum  value of $n$.
Hence, the Hamiltonian in each region $(n)$, defined by $h(n - 1) < x < hn$, is given by
\begin{equation} \label{GrindEQ3}
H_n= p_{x}\sigma_{x}+m\sigma_{z}+V_n.
\end{equation}
Actually we can even generalize this approach to treat spatially dependent masses in which case $m_n=m(x_n)$ in the above equation, however from now on we limit ourselves to a uniform mass distribution. In each spacial region we have a 1D Dirac fermion subject to a constant potential, hence the solution is easily generated. The complete solution for (\ref{GrindEQ3})  satisfying eigenequation with spinor $\psi_n(x)=(\psi^1_n~~\psi^2_n)^t$ in the $n$-th space region where the potential is approximated by it local value $V_n$ will depend on the energy range and can be
written as
%for $\varepsilon\leq V_n-m$:
\begin{eqnarray}
\psi_n(x) &=& A_{n}\left(
                 \begin{array}{c}
                    i \alpha_n \\
                    1\\
                 \end{array}
               \right)e^{i p_{n}x}+B_n\left(
                            \begin{array}{c}
                              -i \alpha_n  \\
                              1\\
                            \end{array}
                          \right)e^{-i p_{n} x}, \qquad \varepsilon\leq V_n-m \label{GrindEQ4}\\
\psi_n(x) &=& A_n \left(
            \begin{array}{c}
              \alpha_n  \\
              1\\
            \end{array}
          \right)e^{ p_{n} x}+B_n\left(
                       \begin{array}{c}
                         - \alpha_n \\
                         1\\
                       \end{array}
                     \right)e^{- p_{n} x}, \qquad V_n-m\leq\varepsilon\leq V_n \label{GrindEQ5}\\
\psi_n(x) &=& A_n\left(
            \begin{array}{c}
              1 \\
              \alpha_n   \\
            \end{array}
          \right)e^{ p_{n} x}+B_n\left(
                       \begin{array}{c}
                        1\\
                        - \alpha_n  \\
                       \end{array}
                     \right)e^{- p_{n} x}, \qquad V_n\leq\varepsilon\leq V_n+m \label{GrindEQ6}\\
\psi_n(x) &=& A_n\left(
               \begin{array}{c}
                 1 \\
                  i \alpha_n \\
               \end{array}
             \right)e^{i p_{n} x} +B_n\left(
                          \begin{array}{c}
                            1\\
                            -i \alpha_n \\
                          \end{array}
                        \right)e^{-i p_{n} x}, \qquad \varepsilon\geq V_n+m \label{GrindEQ7}
\end{eqnarray}
where we have set
\beq
\alpha_{n}=\sqrt{\left|\frac{\left|\varepsilon-V_n\right|-m}{\left|\varepsilon-V_n\right|+m}\right|},
\qquad p_n=\sqrt{\left|(m+V_n-\varepsilon)(m-V_n+\varepsilon)\right|}.
\eeq
The coefficients $A_n$ and $B_n$ are two normalization constants.
The above solutions can be written in a compact matrix form, such as % which reads as follows
\begin{equation} \label{GrindEQ9}
\psi_n(x)=M_n^*(x)\left(
                \begin{array}{c}
                  A_n \\
                  B_n\\
                \end{array}
              \right)
\end{equation}
where the matrix $M_n^*(x)$ is given by
\begin{equation} \label{GrindEQ8}
M_n^*(x)=\left(
         \begin{array}{cc}
            \left[ i^{\frac{1+s^{2}_{n}}{2}}\alpha_{n}\right]^{\left(1-s^{1}_{n}\right)^{\color{red}3}/2} e^{i^{\left(1+s^{1}_{n}\right)/2} p_n x} & \qquad
              \left[- i^{\frac{1+s^{2}_{n}}{2}}\alpha_{n}\right]^{\left(1-s^{1}_{n}\right)^{\color{red}3}/2} e^{-i^{\left(1+s^{1}_{n}\right)/2} p_n x}\\
             \left[ i^{\frac{1+s^{2}_{n}}{2}}\alpha_{n}\right]^{\left(1+s^{1}_{n}\right)/2} e^{i^{\left(1+s^{1}_{n}\right)/2} p_n x} & \qquad
               \left[ -i^{\frac{1+s^{2}_{n}}{2}}\alpha_{n}\right]^{\left(1-s^{1}_{n}\right)^{\color{red}3}/2} e^{-i^{\left(1+s^{1}_{n}\right)/2} p_n x}\\
         \end{array}
       \right).
\end{equation}
 $s^{1}_{n}=\mbox{sign}(\varepsilon-V_n)$ and $s^{2}_{n}=\mbox{sign}(|\varepsilon-V_n|-m)$ are the usual sign functions, equal to $\pm$ for a positive and negative argument, respectively. In order to treat scattering problems,
and for ease of implementation, we consider an incident wave propagating from right to left, this amounts to change $i$ into $-i$ in our previous spinor solution.

\begin{center}
\includegraphics [width=10.10cm,keepaspectratio]
{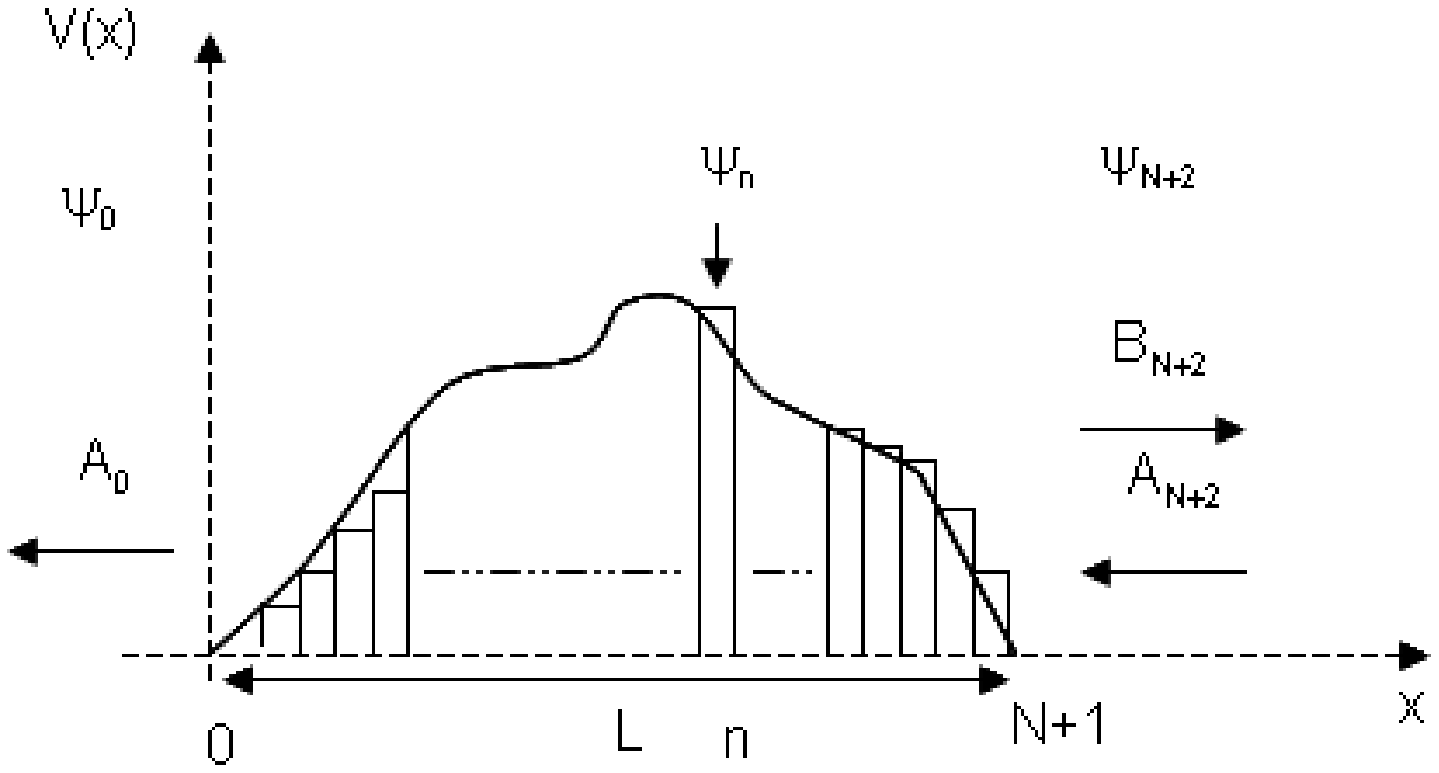}\\
%\end{center}
%\begin{center}
{\sf{Figure 1: Space discretization of the scattering potential $V(x)$ }}\\
\end{center}
%\end{figure}

\begin{center}
\includegraphics [width=10.10cm,keepaspectratio]
{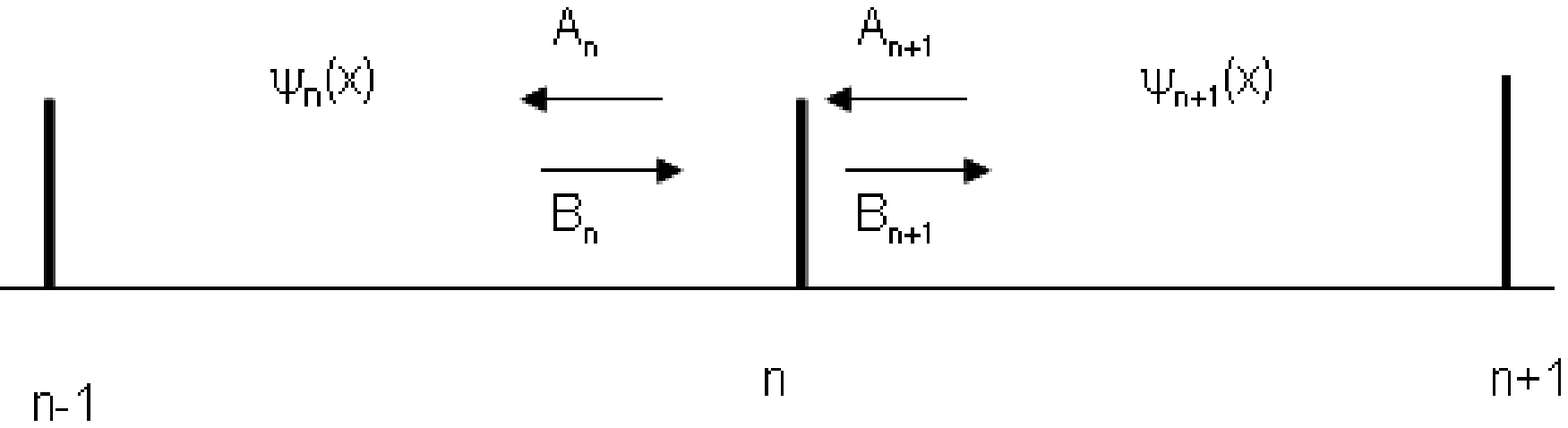}\\
 % Image001.jpg: 354x155 pixel, 96dpi, 9.37x4.10 cm, bb=0 0 266 116
%\caption
%\end{center}
%\begin{center}
{\sf{Figure 2: Solution space for two consecutive regions }}
\end{center}
%\end{figure}

To obtain the desired Poincar\'e Map we %define $\psi_n=\psi_n(x=nh)$,  $M_n = M_n(x=nh)$ and
use the continuity of the spinor wavefunction at the junction $x=x_n$ separating the $n$-th and $(n+1)$-th region. This gives the  relationship
\begin{equation} \label{GrindEQ14}
\begin{array}{cc}
  M_{n}(x_n)\left(
                  \begin{array}{c}
                    A_{n} \\
                    B_{n} \\
                  \end{array}
                \right)=L_{n+1} (x_n)\left(
                  \begin{array}{c}
                    A_{n+1} \\
                    B_{n+1} \\
                  \end{array}
                \right), \qquad  L_{n+1}(x_n)=M_{n+1}(x_n).
\end{array}
\end{equation}
We also have
%We can also write the following
a relationship between $M_{n+1}$ and $L_{n+1}$, which reads
\begin{equation}\label{GrindEQ17}
\begin{array}{cc}
  M_{n+1}(x_{n+1})=L_{n+1}(x_{n}) S_{n+1}, \qquad  S_{n+1}=\left(
         \begin{array}{cc}
            e^{-i^{\left(1+s^{2}_{n+1}\right)/{2}}p_{n+1} h} &
             0\\
           0 &
             e^{i^{\left(1+s^{2}_{n+1}\right)/{2}}p_{n+1} h}\\
         \end{array}
       \right)
\end{array}
\end{equation}
Using the above results we can write the desired Poincar\'e Map as
\begin{equation}\label{GrindEQ21}
\begin{array}{cc}
  \psi_{n+1}(x_{n+1})=\tau_{n}\psi_{n}(x_n), \qquad  \tau _{n}= M_{n+1}(x_{n+1})S_{n+1}M_{n+1}^{-1}(x_{n+1})=
\left(
\begin{array}{cc}
 \tau_{11} &  \tau_{12} \\
  \tau_{21}&  \tau_{22}
\end{array}
\right)
\end{array}
\end{equation}
%\begin{equation}\label{GrindEQ20}
%\tau_{n}=M_{n+1}S_{n+1}M_{n+1}^{-1}
%\end{equation}
where the explicit matrix elements of $\tau_n$ are given by
\begin{equation}
     \tau _{11}=\frac{a(bc-d)}{c-d}, \qquad
\tau _{12}=\frac{a(1-b)}{c-d}, \qquad
\tau _{21}=\frac{a(b-1)c\left(\alpha_{n+1}\right)^{s^{1}_{n+1}}}{c-d}, \qquad
\tau _{22}=\frac{a(c-bd)}{c-d}
\end{equation}
and
\begin{eqnarray}
&&  a = e^{-i^{\left(1+s^{2}_{n+1}\right)/{2}}p_{n+1} h}, \qquad
b=e^{2 i^{\left(1+s^{2}_{n+1}\right)/{2}}p_{n+1} h} \\
&& c = \left(-i^{\left(1+s^{2}_{n+1}\right)/{2} }\alpha_{n+1}\right)^{s^{1}_{n+1}},\qquad
d=\left(i^{\left(1+s^{2}_{n+1}\right)/{2} }\alpha_{n+1}\right)^{s^{1}_{n+1}}.
\end{eqnarray}

Now we proceed to determine the transmission amplitude $t$ using the above recursive relationship, which connects
$A_{n }$ and $\psi_{n}$. For $x\leq 0$ where $V_{0} = 0$ (i.e. left region), we can use for our transmitted spinor
evaluated at $n = 0$, which when suitably normalized can be written as
\begin{equation}\label{GrindEQ10}
    \psi_{0}=\left(
               \begin{array}{c}
                 1 \\
                 -i\alpha_0 \\
               \end{array}
             \right).
\end{equation}
On the other side, for $x\geq L$ were $V_{0}=0$ (i.e. right region), we have both incident and reflected spinor waves. Just outside the potential region on the right hand side in the $(N+2)$-th region the spinor wave is given by
\begin{equation} \label{GrindEQ12}
\psi_{R}(x)=A_{N+2}\left(
               \begin{array}{c}
                 1 \\
                  -i \alpha_0 \\
               \end{array}
             \right)e^{-i p_{0} x} +B_{N+2}\left(
                          \begin{array}{c}
                            1\\
                            i \alpha_0 \\
                          \end{array}
                        \right)e^{i p_{0} x}.
\end{equation}
From the above notation we can easily define the transmission amplitude as
\begin{equation}\label{GrindEQ13}
t=\frac{1}{A_{N+2}}.
\end{equation}
Hence to evaluate the transmission amplitude all we need is to find $A_{N+2}$ using the above recursive scheme.
Our strategy now is to express $A_{N+2}$ in terms of the two end points spinors
$\psi_{N+1}= \psi_{N+1}(x_{N+1})$ and $\psi_{N+2}=\psi_{N+2}(x_{N+2})$.
This can be easily done using our previous relations and leads to the form
\begin{eqnarray}\label{GrindEQ35}
A_{N+2} &=& \frac{e^{i^{\left(1+s^{2}_{0}\right)/2}p_{0} h(N+2)}} {\left(\alpha_0^{2}+1\right)
\left(1-e^{2i^{\left(1+s^{2}_{0}\right)/{2}}p_0 h}\right)}\\
&& \left(
 \begin{array}{ccc}
 \left[ i^{\left(1+s^{2}_{0}\right){2}}\alpha_{0}\right]^{\left({1-s^{1}_{0}}/{2}\right)}&\ \ &
 \left[ i^{\left({1+s^{2}_{0}}\right)/{2}}
 \alpha_{0}\right]^{\left({1+s^{1}_{0}}\right)/2} \\
 \end{array}
 \right)\left(\psi_{N+2}-e^{i^{\left({1+s^{2}_{0}}\right)/{2}} p_0 h}\psi_{N+1}\right)\nonumber.
\end{eqnarray}
Summing up, we iterate the Poinca\'e Map  (\ref{GrindEQ21}) to obtain the end point spinors, $\psi_{N+1}$ and $\psi_{N+2}$,
in terms of the transmitted signor $\psi_0$. These spinors will then be injected in equations (\ref{GrindEQ35}) and   (\ref{GrindEQ13}) to determine the transmission amplitude.

To test the validity of our previous approach we will now implement it for a test potential, the Woods--Saxon
potential which has an exact analytical solution \cite{13, 14}. This potential is defined by \cite{14}
\begin{equation}\label{GrindEQ22}
V(x)=V_0 \left(\frac{\theta (-x)}{1+e^{-a (x+L)}}+\frac{\theta (x)}{1+e^{a (x-L)}}\right)
\end{equation}
where $V_0$ is a real and positive for a barrier or negative for a well, $a$ and $L$ are real and positive. $\theta (x)$ is the Heaviside step function. The analytical solutions provided in reference \cite{14} is used to evaluate the exact transmission coefficient for the potential
(\ref{GrindEQ22}) from the asymptotic behavior of the wavefunctions. This analysis leads to
\begin{equation}
 T=1-\left|\frac{B}{A}\right|^{2}\frac{E+k}{E-k}
\end{equation}
where $A$ and $B$ are given by %the ratio $\frac{B}{A}$ read as
%\begin{equation}
 %\frac{B}{A}= \frac{\alpha_{21}+\beta \alpha_{22}}{\alpha_{11}+\beta \alpha_{12}}
%\end{equation}
%and different parameters are given in terms of Gamma function as
\begin{eqnarray}
    A&=& D_1\frac{\Gamma(1-2\mu) \Gamma(-2\nu)}{\Gamma(-\mu-\nu-\lambda) \Gamma(1-\mu-\nu+\lambda)}e^{-i\pi \mu} +
    D_2\frac{\Gamma(1+2\mu) \Gamma(-2\nu)}{\Gamma(\mu-\nu-\lambda) \Gamma(1+\mu-\nu+\lambda)}e^{i\pi \mu} \nonumber \\
   B &=& D_1\frac{\Gamma(1-2\mu) \Gamma(2\nu)}{\Gamma(-\mu+\nu-\lambda) \Gamma(1-\mu+\nu+\lambda)}e^{-i\pi \mu} +
D_2\frac{\Gamma(1+2\mu) \Gamma(2\nu)}{\Gamma(\mu+\nu-\lambda) \Gamma(1+\mu+\nu+\lambda)}e^{i\pi \mu}
\end{eqnarray}
with the ratio
\beq
      \frac{D_2}{D_1}= \frac{\Gamma(2\mu) \Gamma(1-\mu-\nu-\lambda)\Gamma(-\mu-\nu+\lambda)}{\Gamma(-2\mu) \Gamma(1+\mu-\nu-\lambda)\Gamma(\mu-\nu+\lambda)}e^{-2i\pi \mu}e^{4aL \mu}\nonumber
\eeq
and the abbreviations
     $\mu=\frac{i p}{a},  \nu=\frac{i k}{a},  \lambda=\frac{iV_{0}}{a},$
    $p^{2}=(E-V_{0})^{2}-m^{2}$, $ k^{2}=E^{2}-m^{2}$ have been used.

For scattering states, of interest to us, $|E| > m$ ensures that $k$ is real while the momentum $p$ is real for $m < E < V_0-m$ (the Klein range) and $E>V_0 +m$, it is imaginary for $V_0 -m < E < V_0 + m$. The potential strength $V_{0}$ is
real and positive in our computations. In figure 3 below we show the shape of Woods--Saxon potential for
the parameters $L = 10$, $a = 5$ and $V_0 = 1.2$, the vertical lines represent the discretization of this potential. We shifted
the potential to the right by an amount $L$ for convenience, such a translation does not
affect the physics of the problem.\\
\begin{center}
\includegraphics [width=7.8cm,keepaspectratio]
{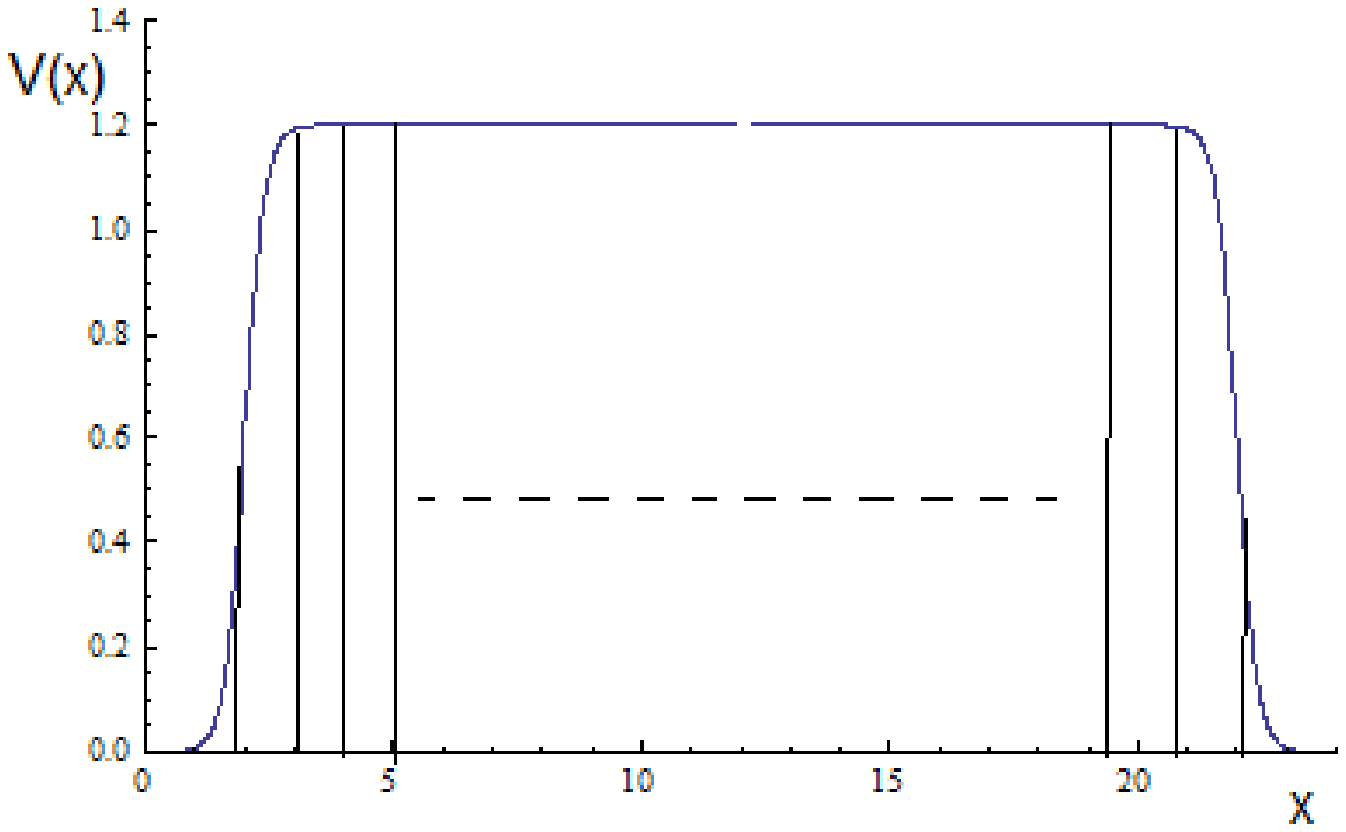}
 % Image001.jpg: 354x155 pixel, 96dpi, 9.37x4.10 cm, bb=0 0 266 116
%\caption
\end{center}
\begin{center}
{\sf{Figure 3: The Woods--Saxon potential curve $V(x)$ shifted to the right by an amount L.
}}
\end{center}
%\end{figure}

\noindent Using the above mentioned numerical procedure (Poincar\'e Map) we evaluate the transmission coefficient
associated with the potential characterized by the above mentioned parameters $(L,a,V_0)$ and a mass $m = 0.4$ in our atomic units.
Moreover,
using the same parameters as in the literature \cite{13}, we have plotted the transmission coefficient in the Klein zone for a constant mass
$m$. As shown in Figure 4 the solid  lines correspond to the exact transmission \cite{13}, and the dashed lines are generated by our Poincar\'e iterative map. We notice from this figure, as expected, that
 as $N$ increases, the dotted line curve  converge to the exact solid line curve. We see the satisfactory agreement between our fast converging numerical approach and the analytical results \cite{13,14}.
\begin{center}
\includegraphics [width=7.7cm,keepaspectratio]
{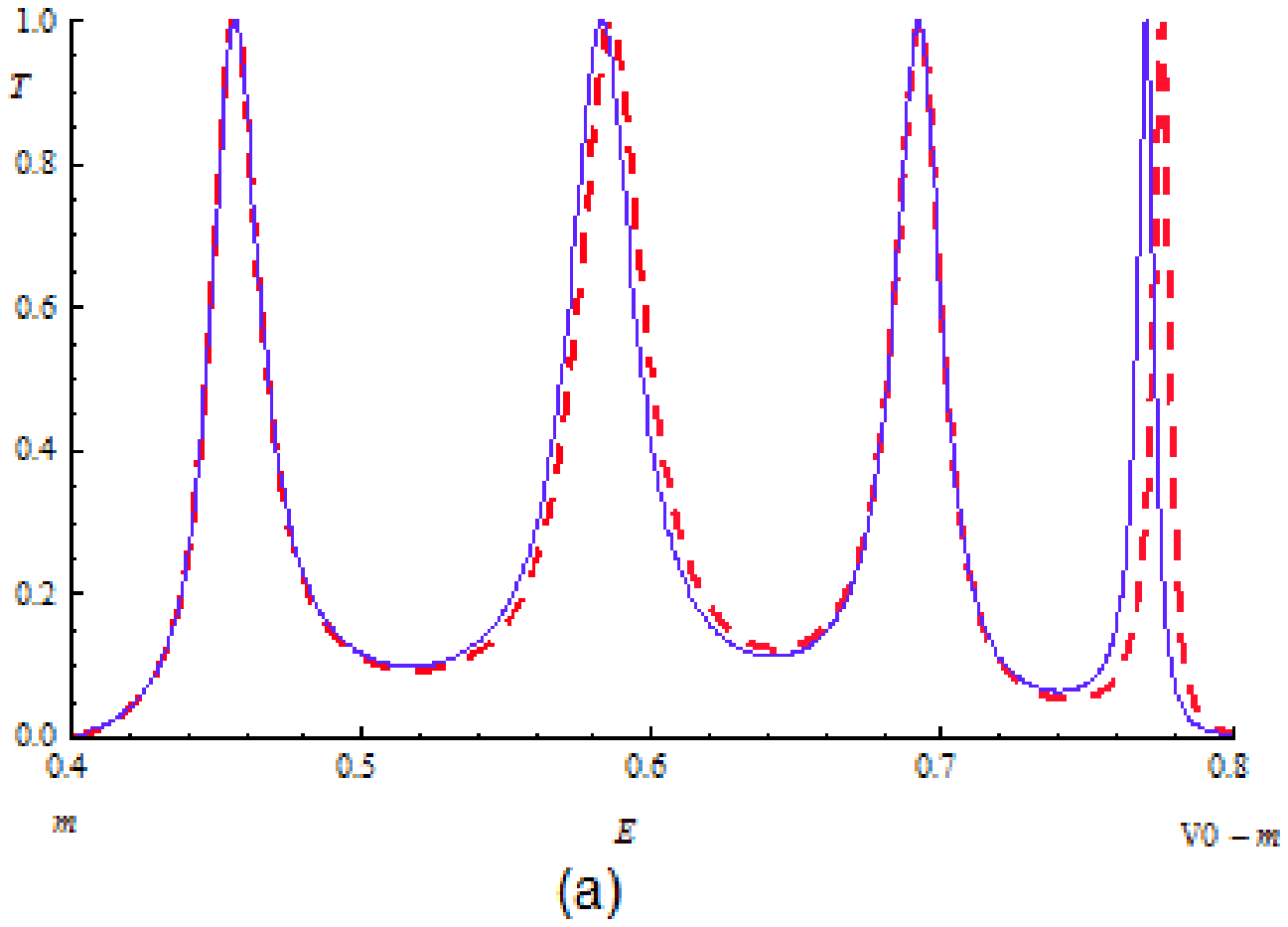}
\includegraphics [width=7.7cm,keepaspectratio]
{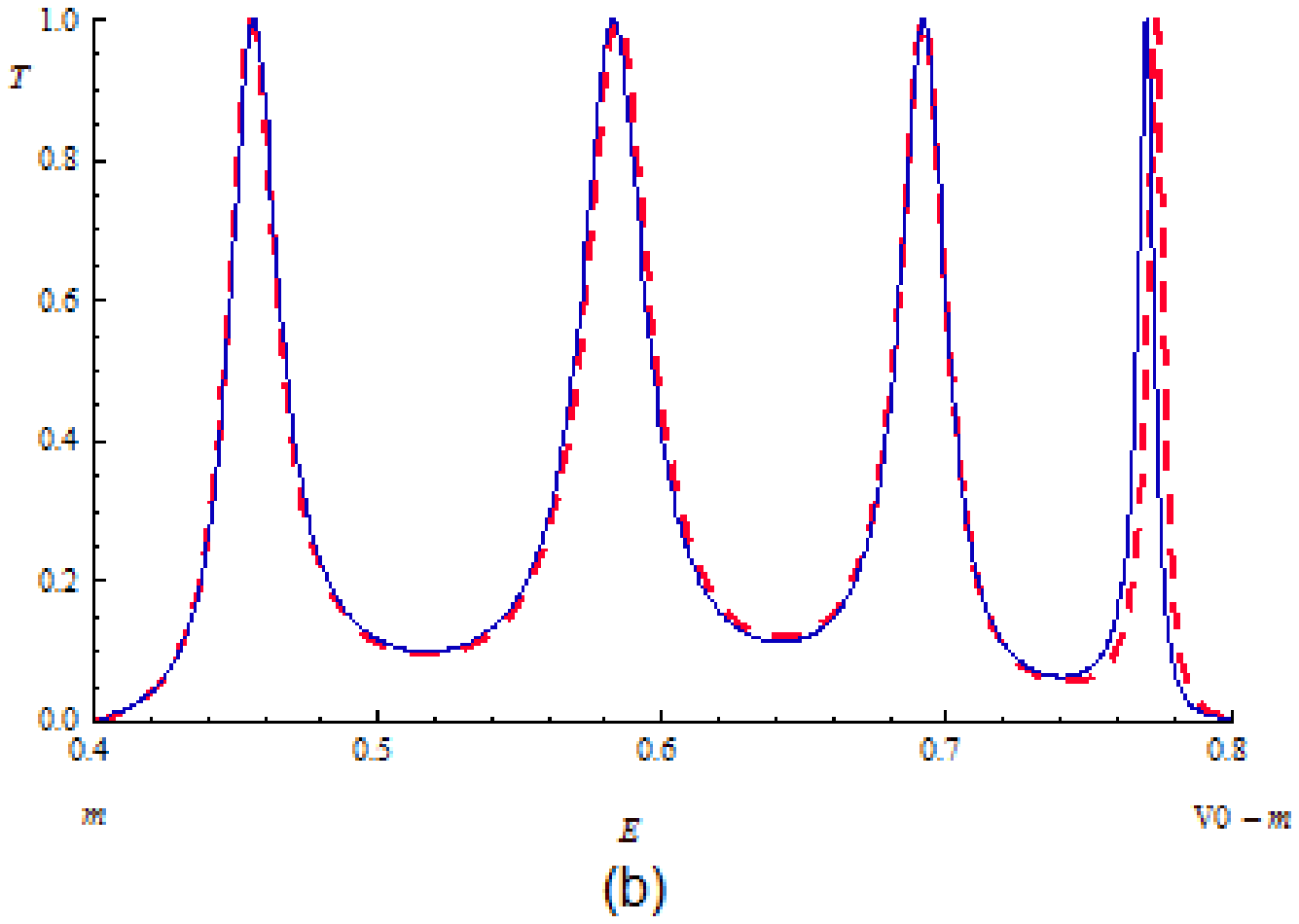}
\end{center}
%\begin{center}
{\sf{Figure 4: The transmission coefficient as a function of energy for two different iterations, (a): $N=600$
and (b): $N=1000$ }}\\

%\end{center}
\noindent In Figure 5 we show the transmission coefficient as a function of the potential strength $V_0$ for two values of the iteration parameter
$N$. The agreement between our numerical approach and the analytical results is impressive. Thus we expect that we can apply our iterative approach to very general potential, which do not lend themselves to analytical solutions. In summary, we believe that the Poincar\'e Map exposed in this work is very simple and enable us to solve the 1D Dirac equation for any arbitrary short range potential with high accuracy and simple computational means. This approach can be easily extended to handle two-dimensional Dirac equation, which is playing an important role in describing the recently discovered graphene system \cite{15}.\\

\begin{center}
\includegraphics [width=7.7cm,keepaspectratio]
{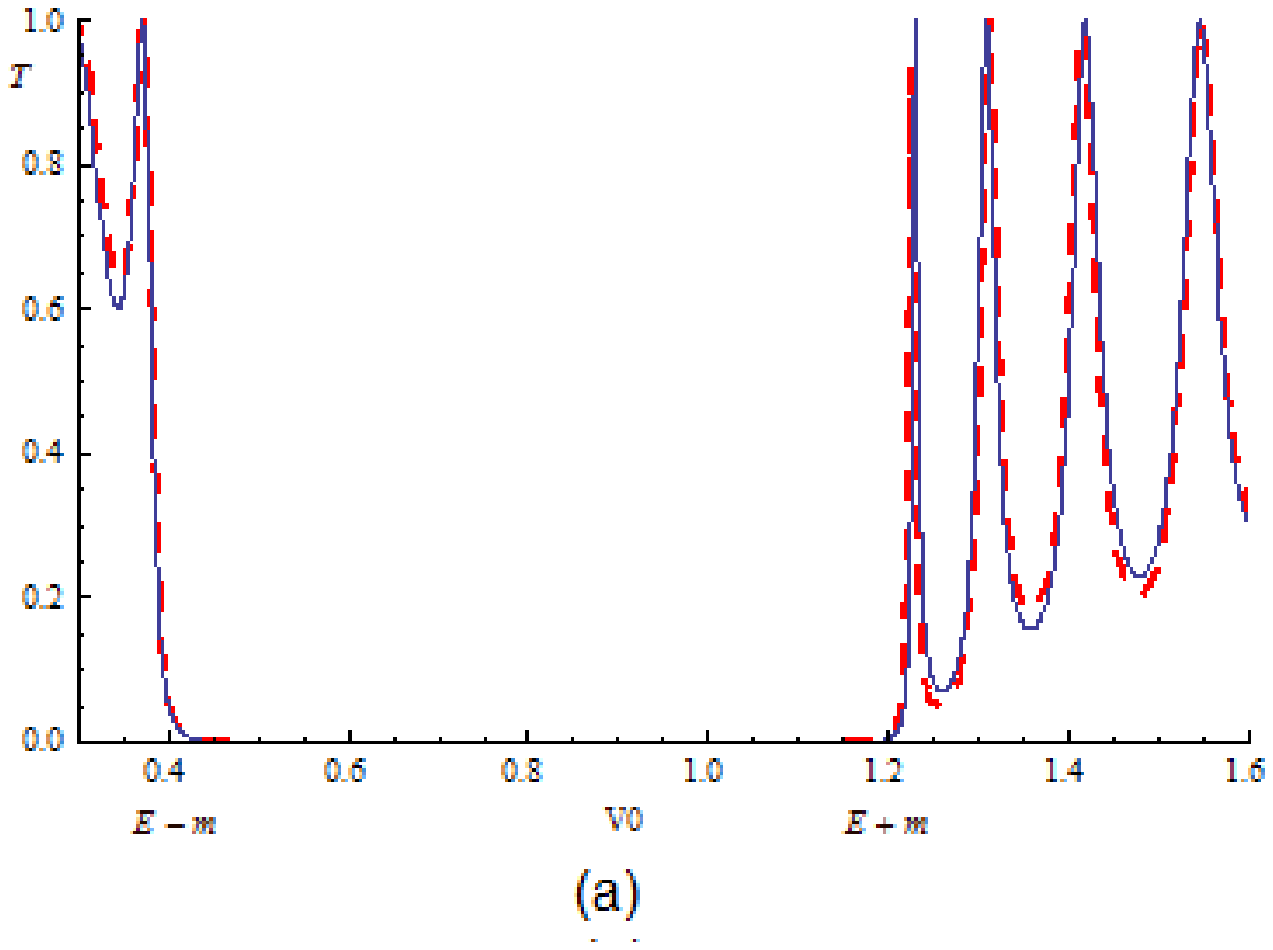}
\includegraphics [width=7.7cm,keepaspectratio]
{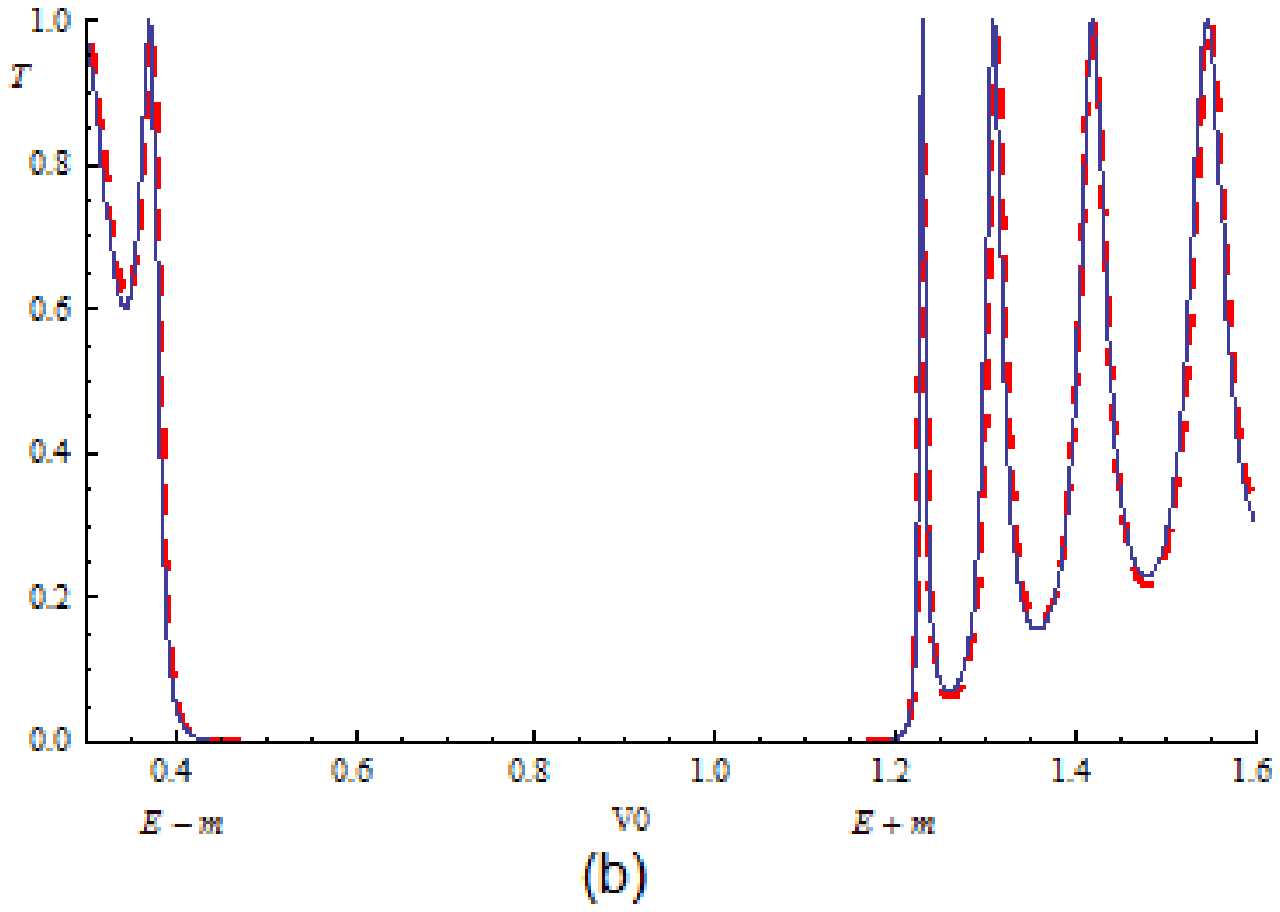}
\end{center}
%\begin{center}
{\sf{Figure 5: The transmission coefficient as a function of $V_0$ for two different iterations, (a): $N=400$ and (b):  $N=1000$}.}
%\end{center}

%%%%%%%%%%%%%%%%%%%%%%%%%%%%%%%%%%%%%%%
\section*{Acknowledgments}
%%%%%%%%%%%%%%%%%%%%%%%%%%%%%%%%%%%%%%%%

The generous support provided by the Saudi Center for Theoretical Physics (SCTP)
is highly appreciated by all Authors. AJ and EBC acknowledge partial support by King Faisal
University and KACST, respectively. We also acknowledge the support of KFUPM under project RG1108-1-2.


\begin{thebibliography}{1}
\bibitem{1} W. Greiner, Relativistic Quantum Mechanics: Wave Equations, (Springer, Berlin, 	1994); J.D. Bjorken and
S.D. Drell, Relativistic Quantum Mechanics, (McGraw-Hill, New York, 1964); W. Greiner, B. Müller and J. Rafelski,
Quantum Electrodynamics of Strong Fields, (Springer, Berlin, 1985).
\bibitem{2} M. Moshinsky and A. Szczepaniak, J. Phys. A 22, L817 (1989).
\bibitem{3} A.D. Alhaidari, Phys. Rev. Lett. 87, 210405 (2002); 88, 189901 (2002).
\bibitem{4} A.D. Alhaidari, J. Phys. A 34, 9827 (2001); 35, 9207 (2002).
\bibitem{5} A.D. Alhaidari, Int. J. Mod. Phys. A 17, 4551 (2002).
\bibitem{6} A.D. Alhaidari, Phys. Rev. A 65, 042109 (2002); 66, 019902 (2002).
\bibitem{7} A.D. Alhaidari, Int. J. Mod. Phys. A 18, 4955 (2003).
\bibitem{8} J.-Y. Guo, X.Z. Fang, and F.-X. Xu, Phys. Rev. A 66, 062105 (2002); J.-Y. Guo,
J. Meng, and F.-X. Xu, Chin. Phys. Lett. 20, 602 (2003).
\bibitem{9} A.D. Alhaidari, Ann. Phys. 312, 144 (2004); Phys. Lett. A 326, 58 (2004); J.
Phys. A 37, 11229 (2004).
\bibitem{10} A.D. Alhaidari, Phys. Lett. A 322, 72 (2004).
\bibitem{11} F. Cooper, A. Khare and Uday Sukhatme, Supersymmetry in Quantum Mechanics, (World Scientific
Publishing Co, 2001).
\bibitem{12} J. Bellisard, A. Formoso, R. Lima and D. Testard, Phys. Rev. B 26, 3024 (1982); F. Dominguez-Adame, E.
Marcia and A. Sanchez, Phys. Rev. B 48, 6054 (1993); E. Diez, A. Sanchez and F. Dominguez-Adame, Phys. Rev. B 50, 14359 (1994).
\bibitem{13} O. Panella, S. Biondini, and A. Arda, J. Phys. A: Math. Theor. {43},  325302 (2010).
\bibitem{14} P. Kennedy, J. Phys. A: Math. Gen. {35},  689 (2002).
\bibitem{15}  A.K. Geim and K.S.  Novoselov,  Nature Mater. {6}, 183 (2007).

\end{thebibliography}
\end{document}